\documentclass{mem}
\usepackage{natbib}\usepackage{txfonts}\usepackage{balance}
\usepackage{graphicx}
\usepackage[a4paper]{hyperref}
\idline{75}{282}
\begin{document}
\def\teff{$T\rm_{eff }$}
\def\kms{$\mathrm {km s}^{-1}$}

\title{
Linking core-collapse supernova explosions to supernova remnants
through 3D MHD modeling
}

   \subtitle{}

\author{
S. \,Orlando\inst{1} 
\and A. \, Wongwathanarat\inst{2}
\and H.-T. \, Janka\inst{2}
\and M. \, Miceli\inst{1,3}
\and M. \, Ono\inst{4}
\and S. \, Nagataki\inst{4}
\and F. \, Bocchino\inst{1}
\and G. \, Peres\inst{1,3}
          }

  \offprints{S. Orlando}

\institute{
INAF -- Osservatorio Astronomico di Palermo, Piazza del Parlamento 1, Palermo, Italy
\email{salvatore.orlando@inaf.it}
\and
Max-Planck-Institut f\"ur Astrophysik, Karl-Schwarzschild-Str. 1, Garching, Germany
\and
Dip. di Fisica e Chimica, Universit\`a di Palermo, Piazza del Parlamento 1, Palermo, Italy
\and
ABBL, RIKEN Cluster for Pioneering Research, 2-1 Hirosawa, Wako, Saitama, Japan
\and
RIKEN, iTHEMS Program, 2-1 Hirosawa, Wako, Saitama, Japan
}

\authorrunning{Orlando }

\titlerunning{Linking core-collapse supernova explosions to supernova
remnants}

\abstract{
The structure and morphology of supernova remnants (SNRs) reflect
the properties of the parent supernovae (SNe) and the characteristics
of the inhomogeneous environments through which the remnants expand.
Linking the morphology of SNRs to anisotropies developed in their
parent SNe can be essential to obtain key information on many aspects
of the explosion processes associated with SNe. Nowadays, our
capability to study the SN-SNR connection has been largely improved
thanks to multi-dimensional models describing the long-term evolution
from the SN to the SNR as well as to observational data of growing
quality and quantity across the electromagnetic spectrum which allow
to constrain the models. Here we used the numerical resources
obtained in the framework of the ``Accordo Quadro INAF-CINECA
(2017)'' together with a CINECA ISCRA Award N.HP10BARP6Y to describe
the full evolution of a SNR from the core-collapse to the full-fledged
SNR at the age of 2000 years. Our simulations were compared
with observations of SNR Cassiopeia A (Cas A) at the age of $\sim
350$~years. Thanks to these simulations we were able to link
the physical, chemical and morphological properties
of a SNR to the physical processes governing the complex phases
of the SN explosion.

\keywords{hydrodynamics –- instabilities –- shock waves –- ISM: supernova
remnants –- X-rays: ISM –- supernovae: individual (Cassiopeia A) }

}
\maketitle{}

\section{Introduction}

Supernova remnants (SNRs), the outcome of supernova (SN) explosions,
are diffuse extended sources with a complex morphology and a highly
non-uniform distribution of ejecta. General consensus is that such
morphology reflects, on one hand, the physical and chemical properties
of the parent SN and, on the other hand, the properties of the
progenitor stars and the early interaction of the SN blast wave
with the circumstellar medium (CSM). Thus investigating the link
between the morphological properties of a SNR and the complex phases
in the SN explosion may help: 1) to trace back the characteristics
of the asymmetries that may have occurred during the SN explosion,
providing a physical insight into the processes governing the SN
engines; 2) to identify the imprint of the progenitor stars,
thereby obtaining information on the final stages of the stellar
evolution.

However, studying the connection between SNRs and their parent SNe
is an extremely challenging task which requires a multi-physics,
multi-scale, multi-dimensional approach, due to the rich physics involved
during the SN explosion and the subsequent expansion of the remnant,
the very different time and space scales involved in these phases
of evolution, and the inherent 3D nature of the anisotropies
developing during the whole evolution. To overcome all these
difficulties, we adopted an approach based on the coupling between
core-collapse SN models and SNR models (e.g.  \citealt{2015ApJ...810..168O,
2016ApJ...822...22O, 2017ApJ...842...13W,
2019ApJ...877..136F, 2020ApJ...888..111O, 2020A&A...636A..22O,
2020arXiv200901157T, 2020arXiv200801763G}).

Here we studied the ejecta dynamics of the SNR Cas A from the
core-collapse SN explosion to their expansion in the SNR by linking,
for the first time, modeling attempts that have been carried out
independently so far, either constrained to the early phase of the
SN up to days only (\citealt{2017ApJ...842...13W}), or starting the
long-time remnant evolution with artificial initial conditions
(\citealt{2016ApJ...822...22O}). The simulations covered 2000~years
of evolution, thus going beyond the age of Cas A ($\approx 350$~years).
The aim was to investigate how the final remnant morphology reflects
the characteristics of the SN explosion and, in particular, the
asymmetries that develop in the immediate aftermath of the core-collapse

\section{The model}

We focused on a SN model which reproduces post-explosion anisotropies,
one day after the SN event, which are compatible with the structure
of Cas A (\citealt{2017ApJ...842...13W}): three pronounced nickel-rich
fingers that may correspond to the extended iron-rich regions
observed in Cas A. These simulations provided the initial conditions
for our 3D SNR simulations soon after the shock breakout. 
The plasma and magnetic field evolution were modeled numerically
by solving the time-dependent MHD equations, including the deviations
from electron-proton temperature-equilibration, and the deviations
from equilibrium of ionization of the most abundant ions, in a 3D
Cartesian coordinate system (see \citealt{2020arXiv200901789O}
for the detail of the implementation). These effects are necessary
to describe appropriately the evolution of the remnant and to
synthesize accurately the thermal X-ray emission from model results.
We also included the effects of heating due to radioactive decay
of $^{56}$Ni and $^{56}$Co.

The calculations were performed using the PLUTO code
(\citealt{2007ApJS..170..228M}) a well tested modular Godunov-type
code intended mainly for astrophysical applications and high Mach
number flows in multiple spatial dimensions. The initial density,
pressure, and velocity structure of ejecta as well as the isotopic
composition of the ejecta were derived from the adopted SN simulation
(\citealt{2017ApJ...842...13W}). Then the 3D SNR simulations describe
the interaction of the remnant with the wind of the progenitor star.
The wind was assumed to be spherically symmetric with gas density
proportional to $r^{-2}$ (where $r$ is the radial distance from the
progenitor). In order to compare the simulations with Cas A, the
wind density at $r = 2.5$~pc was constrained by X-ray observations
of the shocked wind in this remnant (\citealt{2014ApJ...789....7L}).

We performed multi-species simulations to follow the evolution of
the isotopic composition of ejecta and the matter mixing and to
link the chemical distribution of ejecta in the remnant to anisotropies
developing in the early phases of SN evolution.  A major challenge
in modeling the explosion and subsequent evolution of the remnant
was the very small scale of the system (the initial blast wave
radius is $\approx 10^{14}$~cm) in the immediate aftermath of the
SN explosion ($\approx 1$~day after the SN event) in comparison
with the size of the rapidly expanding blast wave (a final size of
$\approx 8$~pc at the age of 2000~years). To capture this
range of scales we adopted a strategy similar to that used by
\cite{2019A&A...622A..73O}.

\section{Results}

The model and the results of our analysis are presented in detail in
\cite{2020arXiv200901789O}; here we summarize our main findings.

We explored the effects of energy deposition from radioactive decay
and the effects of an ambient magnetic field on the evolution of
the remnant, by performing three long-term simulations
with the above effects switched either on or off. In all the cases,
our simulations predict radii of the forward and reverse shocks and
ejecta velocities at the age of $\approx 350$~years consistent with
those observed in Cas A. We found that the fundamental chemical,
physical, and geometric properties observed in Cas A can naturally
be explained in terms of the physical processes associated with the
asymmetric SN explosion and to subsequent interaction of the initial
asymmetries with the reverse shock of the SNR.
\begin{figure}[t!]
\resizebox{\hsize}{!}{\includegraphics[clip=true]{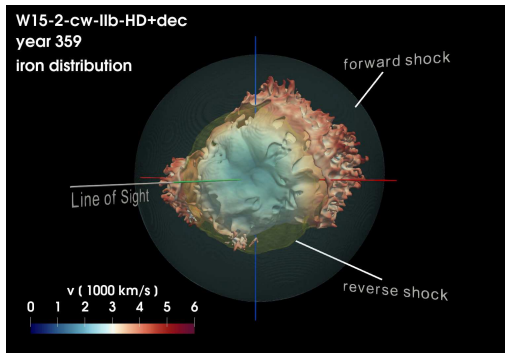}}
\caption{\footnotesize Isosurface of the distribution of iron at
the age of Cas A for one of our simulations (see
\citealt{2020arXiv200901789O}). The opaque isosurface correspond
to a value of Fe density which is at 5\% of the peak density; the
colors give the radial velocity in units of 1000 km s$^{-1}$ on the
isosurface. The semi-transparent quasi-spherical surfaces indicate
the forward (green) and reverse (yellow) shocks. The Earth vantage
point lies on the negative $y$-axis. A navigable 3D graphic of this
distribution is available at https://skfb.ly/6TKRK.
}
\label{fig1}
\end{figure}

The ejecta distribution soon after the breakout of the shock wave
at the stellar surface is characterized by large-scale plumes of
ejecta rich in $^{56}$Ni and $^{44}$Ti (\citealt{2017ApJ...842...13W}).
One year later, the $^{56}$Ni decayed in $^{56}$Co and the latter
in stable $^{56}$Fe.  The high-entropy plumes of Fe- and Ti-rich
ejecta cross the reverse shock at the age of $\approx 30$~years.
This interaction triggers the development of hydrodynamic (HD)
instabilities which gradually fragment the plumes into numerous
small-scale fingers. At the age of Cas A, the shocked portion of
these plumes lead to the formation of extended regions of shock-heated
Fe similar to those observed in Cas A. The fine-scale structure of
the plumes and the HD instabilities triggered by the reverse shock
lead to the formation of a filamentary pattern of shocked ejecta
with ring-like and crown-like features (see Fig.~\ref{fig1}) which
resembles that observed in Cas A remarkably well
(e.g.~\citealt{2013ApJ...772..134M}). The initial large-scale plumes
are also responsible for the spatial inversion of the ejecta layers
with Si-rich ejecta being physically interior to Fe-rich ejecta.
The inversion is evident only in regions where the fast plumes of
Fe-rich ejecta interact with the reverse shock; elsewhere, the
original chemical stratification is roughly preserved.

\begin{figure}[t!]
\resizebox{\hsize}{!}{\includegraphics[clip=true]{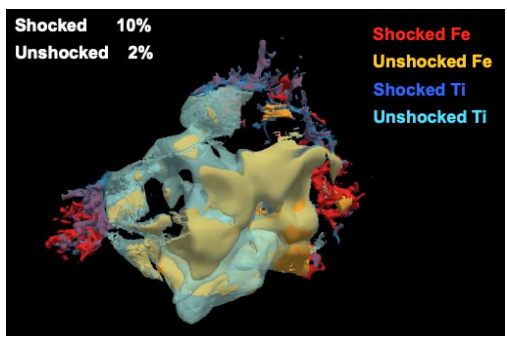}}
\caption{\footnotesize Isosurfaces of the distribution of iron and
titanium at the age of Cas A for one of our simulations (see
\citealt{2020arXiv200901789O}). The irregular isosurfaces correspond
to a value of Fe and Ti density at 10\% (shocked) or 2\% (unshocked)
of the maximum density; the colors indicate either shocked or
unshocked Fe and Ti as reported in the upper right corner.
}
\label{fig2}
\end{figure}

The unshocked ejecta of the modeled remnant show voids and cavities
similar to those observed (e.g.~\citealt{2015Sci...347..526M}; see
Fig.~\ref{fig2}).  They originate from the expansion of Fe-rich
plumes and their inflation due to the decay of radioactive species.
We found that the largest cavities are physically connected with
the large-scale Si-rich rings of shocked ejecta which encircle the
Fe-rich regions.  The initial large-scale asymmetry is also responsible
for the metal-rich ejecta being arranged in a ``thick-disk'' geometry.
The disk is tilted with respect to the plane of the sky as inferred
from observations of Cas A. The distributions of $^{44}$Ti and
$^{56}$Fe are mostly concentrated in the northern hemisphere,
pointing opposite to the kick velocity of the neutron star. These
distributions and their abundance ratio are both compatible with
those inferred from high-energy observations of Chandra and NuSTAR
(\citealt{2014Natur.506..339G, 2017ApJ...834...19G}).

Finally, the simulations showed that, after 2000~years of evolution,
most of the metal-rich ejecta were shocked and subject to strong
mixing by the HD instabilities. However, the spatial distributions
of iron-group elements, the decay product of Ti ($^{44}$Ca) and Si
still keep memory of the original SN asymmetry. Furthermore, cavities
and voids in the unshocked ejecta are still evident even though
more diluted than in previous epochs. We concluded that the
fingerprints of the SN can still be found $\approx 2000$~years
after the explosion.

\section{Conclusions}

Our study has shown that the structures observed in the inter-shock
region of a core-collapse SNR are the natural consequence of the
interaction of the reverse shock with the post-explosion large-scale
asymmetries of the SN. In particular, the main features observed
in the morphology of Cas A (Fe-rich regions, ring- and crown-like
features, inversion of ejecta layers, etc.) develop as a consequence
of the fast growth of HD instabilities in the reverse-shock heated
plumes of ejecta developed after the SN explosion. All these features
encode the fingerprints of a neutrino-driven SN explosion. Our
simulations, therefore, provided more conclusive information
whether the paradigm of the neutrino-driven explosion mechanism is
able to explain the chemical and morphological asymmetries observed
in great detail in Cas A, a young remnant of a stripped-envelope
(type IIb) SN.

\begin{acknowledgements}
The PLUTO code is developed at the Turin Astronomical Observatory
(Italy) in collaboration with the Department of General Physics of
Turin University (Italy) and the SCAI Department of CINECA (Italy).
We acknowledge the ``Accordo Quadro INAF-CINECA (2017)'' and CINECA
ISCRA Award N.HP10BARP6Y for the availability of high performance
computing resources and support at the infrastructure Marconi based
in Italy at CINECA. At Garching, funding through ERC-AdG no.\
341157-COCO2CASA and DFG grants SFB-1258 and Cluster of Excellence
ORIGINS (EXC-2094)390783311 is acknowledged. Computer resources
have been provided by MPCDF.

\end{acknowledgements}

\bibliographystyle{aa}
\bibliography{references}

\begin{thebibliography}{16}
\expandafter\ifx\csname natexlab\endcsname\relax\def\natexlab#1{#1}\fi

\bibitem[{{Ferrand} {et~al.}(2019){Ferrand}, {Warren}, {Ono}, {Nagataki},
  {R{\"o}pke}, \& {Seitenzahl}}]{2019ApJ...877..136F}
{Ferrand}, G., {Warren}, D.~C., {Ono}, M., {et~al.} 2019, \apj, 877, 136

\bibitem[{{Gabler} {et~al.}(2020){Gabler}, {Wongwathanarat}, \&
  {Janka}}]{2020arXiv200801763G}
{Gabler}, M., {Wongwathanarat}, A., \& {Janka}, H.-T. 2020, arXiv e-prints,
  arXiv:2008.01763

\bibitem[{{Grefenstette} {et~al.}(2017){Grefenstette}, {Fryer}, {Harrison},
  {Boggs}, {DeLaney}, {Laming}, {Reynolds}, {Alexander}, {Barret},
  {Christensen}, {Craig}, {Forster}, {Giommi}, {Hailey}, {Hornstrup},
  {Kitaguchi}, {Koglin}, {Lopez}, {Mao}, {Madsen}, {Miyasaka}, {Mori}, {Perri},
  {Pivovaroff}, {Puccetti}, {Rana}, {Stern}, {Westergaard}, {Wik}, {Zhang}, \&
  {Zoglauer}}]{2017ApJ...834...19G}
{Grefenstette}, B.~W., {Fryer}, C.~L., {Harrison}, F.~A., {et~al.} 2017, \apj,
  834, 19

\bibitem[{{Grefenstette} {et~al.}(2014){Grefenstette}, {Harrison}, {Boggs},
  {Reynolds}, {Fryer}, {Madsen}, {Wik}, {Zoglauer}, {Ellinger}, {Alexand er},
  {An}, {Barret}, {Christensen}, {Craig}, {Forster}, {Giommi}, {Hailey},
  {Hornstrup}, {Kaspi}, {Kitaguchi}, {Koglin}, {Mao}, {Miyasaka}, {Mori},
  {Perri}, {Pivovaroff}, {Puccetti}, {Rana}, {Stern}, {Westergaard}, \&
  {Zhang}}]{2014Natur.506..339G}
{Grefenstette}, B.~W., {Harrison}, F.~A., {Boggs}, S.~E., {et~al.} 2014, \nat,
  506, 339

\bibitem[{{Lee} {et~al.}(2014){Lee}, {Park}, {Hughes}, \&
  {Slane}}]{2014ApJ...789....7L}
{Lee}, J.-J., {Park}, S., {Hughes}, J.~P., \& {Slane}, P.~O. 2014, \apj, 789, 7

\bibitem[{{Mignone} {et~al.}(2007){Mignone}, {Bodo}, {Massaglia}, {Matsakos},
  {Tesileanu}, {Zanni}, \& {Ferrari}}]{2007ApJS..170..228M}
{Mignone}, A., {Bodo}, G., {Massaglia}, S., {et~al.} 2007, \apjs, 170, 228

\bibitem[{{Milisavljevic} \& {Fesen}(2013)}]{2013ApJ...772..134M}
{Milisavljevic}, D. \& {Fesen}, R.~A. 2013, \apj, 772, 134

\bibitem[{{Milisavljevic} \& {Fesen}(2015)}]{2015Sci...347..526M}
{Milisavljevic}, D. \& {Fesen}, R.~A. 2015, Science, 347, 526

\bibitem[{{Ono} {et~al.}(2020){Ono}, {Nagataki}, {Ferrand}, {Takahashi},
  {Umeda}, {Yoshida}, {Orland o}, \& {Miceli}}]{2020ApJ...888..111O}
{Ono}, M., {Nagataki}, S., {Ferrand}, G., {et~al.} 2020, \apj, 888, 111

\bibitem[{{Orlando} {et~al.}(2019){Orlando}, {Miceli}, {Petruk}, {Ono},
  {Nagataki}, {Aloy}, {Mimica}, {Lee}, {Bocchino}, {Peres}, \&
  {Guarrasi}}]{2019A&A...622A..73O}
{Orlando}, S., {Miceli}, M., {Petruk}, O., {et~al.} 2019, \aap, 622, A73

\bibitem[{{Orlando} {et~al.}(2015){Orlando}, {Miceli}, {Pumo}, \&
  {Bocchino}}]{2015ApJ...810..168O}
{Orlando}, S., {Miceli}, M., {Pumo}, M.~L., \& {Bocchino}, F. 2015, \apj, 810,
  168

\bibitem[{{Orlando} {et~al.}(2016){Orlando}, {Miceli}, {Pumo}, \&
  {Bocchino}}]{2016ApJ...822...22O}
{Orlando}, S., {Miceli}, M., {Pumo}, M.~L., \& {Bocchino}, F. 2016, \apj, 822,
  22

\bibitem[{{Orlando} {et~al.}(2020{\natexlab{a}}){Orlando}, {Ono}, {Nagataki},
  {Miceli}, {Umeda}, {Ferrand}, {Bocchino}, {Petruk}, {Peres}, {Takahashi}, \&
  {Yoshida}}]{2020A&A...636A..22O}
{Orlando}, S., {Ono}, M., {Nagataki}, S., {et~al.} 2020{\natexlab{a}}, \aap,
  636, A22

\bibitem[{{Orlando} {et~al.}(2020{\natexlab{b}}){Orlando}, {Wongwathanarat},
  {Janka}, {Miceli}, {Ono}, {Nagataki}, {Bocchino}, \&
  {Peres}}]{2020arXiv200901789O}
{Orlando}, S., {Wongwathanarat}, A., {Janka}, H.~T., {et~al.}
  2020{\natexlab{b}}, arXiv e-prints, arXiv:2009.01789

\bibitem[{{Tutone} {et~al.}(2020){Tutone}, {Orlando}, {Miceli}, {Ustamujic},
  {Ono}, {Nagataki}, {Ferrand}, {Greco}, {Peres}, \&
  {Warren}}]{2020arXiv200901157T}
{Tutone}, A., {Orlando}, S., {Miceli}, M., {et~al.} 2020, arXiv e-prints,
  arXiv:2009.01157

\bibitem[{{Wongwathanarat} {et~al.}(2017){Wongwathanarat}, {Janka},
  {M{\"u}ller}, {Pllumbi}, \& {Wanajo}}]{2017ApJ...842...13W}
{Wongwathanarat}, A., {Janka}, H.-T., {M{\"u}ller}, E., {Pllumbi}, E., \&
  {Wanajo}, S. 2017, \apj, 842, 13

\end{thebibliography}

\end{document}